\documentclass[fleqn,usenatbib]{mnras}

\usepackage{newtxtext,newtxmath}

\usepackage[T1]{fontenc}

\DeclareRobustCommand{\VAN}[3]{#2}
\let\VANthebibliography\thebibliography
\def\thebibliography{\DeclareRobustCommand{\VAN}[3]{##3}\VANthebibliography}

\usepackage{graphicx}	
\usepackage{amsmath}	

\title[Snowline pebble flux on planet formation]{Simulating the interplay between the snowline pebble flux and ongoing planet formation and migration}

\author[D.~Astrakhantsev et al.]{
Danila Astrakhantsev,$^{1}$\thanks{E-mail: danila.astrakhan@gmail.com}
Sebastiaan Krijt,$^{1}$
Sofia Savvidou,$^{2,3}$ 
Bertram Bitsch$^{4}$
\\
$^{1}$School of Physics and Astronomy, University of Exeter, Stocker Road, Exeter EX4 4QL, UK\\
$^{2}$Space Telescope Science Institute, 3700 San Martin Drive, Baltimore, MD 21218, USA \\
$^{3}$Max Planck Institute for Astronomy, Königstuhl 17, D-69117 Heidelberg, Germany \\
$^{4}$Department of Physics, University College Cork, Cork, T12 R229, Ireland
}

\date{Accepted XXX. Received YYY; in original form ZZZ}

\pubyear{\the\year{}}

\begin{document}
\label{firstpage}
\pagerange{\pageref{firstpage}--\pageref{lastpage}}
\maketitle

\begin{abstract}
Pebble drift plays a central role in modern planet formation models. In this work we carry out planet formation simulations (including pebble accretion and migration) for a range of disc parameters to investigate (a) the impact of the snowline pebble mass flux on final planet orbits and masses, and (b) the back-reaction of growing and migrating planets on the snowline pebble fluxes in their natal discs. We find a strong correlation between the snowline pebble flux (at the time of protoplanet insertion) and the final planet mass. The correlation is continuous in disks with high turbulence levels ($\alpha=10^{-3}$), but exhibits a step function at lower turbulence ($\alpha=10^{-4}$), with giant planet formation requiring (initial) snowline pebble mass fluxes exceeding $100~\mathrm{M_\oplus Myr^{-1}}$. We find qualitative agreement between pebble mass fluxes inferred for discs aged ${\sim}1~\mathrm{Myr}$ and our planet-containing models, especially for larger disks ($\geq$40~au), high $\alpha$ ($10^{-3}$), and low $v_\mathrm{frag}$ ($3\mathrm{~m~s}^{-1}$). Additionally, giant planets in high turbulence disks are found to perturb the snowline pebble flux only temporarily (for ${\approx}10^{5-6}\mathrm{~yr}$) due to them quickly growing and migrating across the snowline. Our simulations show that currently observed pebble fluxes can indeed be used to constrain planet formation simulations, emphasizing that planet formation via pebble accretion is broadly in agreement with the currently available constraints from disc evolution as provided by JWST.
\end{abstract}

\begin{keywords}
protoplanetary discs -- planet–disc interactions -- planets and satellites: formation -- planets and satellites: dynamical evolution and stability
\end{keywords}



\section{Introduction}
Modern planet formation theory reserves a big role for pebble drift, with the large-scale inward migration of marginally-decoupled pebbles featuring centrally in models of planetesimal formation \citep{drazkowska2017, schoonenberg2017, drazkowska_pp7}, planet growth and accretion \citep{johansen2017, ormel2018}, early solar system evolution \citep{lichtenberg2021}, and evolving disc chemistry \citep[e.g.,][]{booth2017, krijt2020, mah_2024, williams2025}. At the same time, qualitative observational constraints on the timing and magnitude of pebble drift have been challenging to obtain, with measurements of e.g., dust disc mass or size evolution challenging to interpret \citep{pascucci2016, najita2018}.

In this context there has been push in recent years to attempt to estimate pebble fluxes at specific locations in discs (i.e., at major snowlines, primarily those of CO, CO$_2$, and H$_2$O) by measuring variations in the gas-phase abundance of major carbon and oxygen carriers -- and associated/attributing those changes to pebble drift and ice sublimation \citep[see][]{bosman2018,zhang2019, zhang2020, Williams_Krijt_2025}. For water in particular, system-to-system variations seen via infrared spectroscopy first with Spitzer and now with the James Webb Space Telescope are frequently attributed to variations in the icy pebble mass capable of reaching the inner disc \citep{najita2013, banzatti2020, banzatti2023, Romero-Mirza_2024, sellek2025}.

Most recently, \citet{Krijt_Banzatti_2025} combined JWST MIRI/MRS and ALMA continuum data for twenty-one protoplanetary discs to (a) measure pebble mass fluxes across the water snowline (building on the method outlined in \citealt{Romero-Mirza_2024}) and (b) demonstrate how those fluxes are shaped by the influence of gaps acting as pebble traps in the outer disc -- a process studied in various numerical works \citep[e.g.,][]{kalyaan_2023, Bitsch_Mah_2023, stammler_2023, easterwood_2024, mah_2024}. Pebble mass flux estimates ranged from ${\sim}1 {-} 10^{3}~M_\oplus/\mathrm{Myr}$ in these ${\sim}\mathrm{Myr}$-old systems, broadly consistent with ranges estimated for the early solar system and values assumed in planet formation models \citep[][Fig.~2]{Krijt_Banzatti_2025, lichtenberg2021}. Given that models indicate that the emerging masses (and architectures) of inner planetary systems are very sensitive to the pebble mass flux arriving in the inner disc \citep[e.g.,][]{lambrechts2019, drazkowska2021, mccloat2025, danti2025}, what does this broad range of inferred snowline pebble fluxes mean for the planets growing in the inner regions of these systems? The aims of this study then are two-fold: (a) study what planet formation models predict in terms of final planet masses as a function of snowline pebble flux; and, at the same time (b) study how accreting and gap-opening planets themselves may in turn affect the snowline pebble mass flux.

This paper is structured as follows: in Sect. \ref{sec:methods} we outline the \texttt{chemcomp} model, how we use it, and the parameter space we explore. We then explore how the snowline pebble flux (measured at either the time of protoplanet insertion or 1 Myr) affects final planet properties in Sects. \ref{sec:t0_fluxes} and Sect. \ref{sec:1Myr_fluxes}. The feedback of growing planets' on the snowline pebble flux is described in Sect. \ref{sec:fluxes_with_planets}. Finally, we discuss how our results connect to JWST observations in Sect. \ref{sec:link_to_JWST}, before highlighting possible extensions and (Sect.~\ref{sec:limitations}) and conclusions.

\section{Methodology}
\label{sec:methods}

\subsection{Model}\label{sec:model}
We use \texttt{chemcomp}\footnote{\url{https://github.com/AaronDavidSchneider/chemcomp}} \citep[][]{Schneider_Bitsch_2021} to study the combined evolution of discs, their dust/pebbles, and planet accretion and migration. This is a semi-analytical 1D model which evolves a viscous protoplanetary disk, including pebble drift, evaporation and condensation, planetary growth and migration, and multiple chemical species. The midplane temperature of the disk is set at the start and is calculated from viscous heating and irradiation from the star. 

The \texttt{chemcomp} code is described in detail in \citet{Schneider_Bitsch_2021, schneider2021b} and has in recent years been used to study, for example giant planet formation \citep[][]{savvidou_bitsch_2023}, the transport and thermal decomposition of refractory organics \citep[][]{houge2025} and the impact of volatile entrapment on evolving C/O ratios \citep{williams2025}. In the remainder of this section we discuss the disc parameters we use, how the snowline pebble mass flux is obtained in our modelling framework, and finally how planets are inserted and evolve using a few illustrative examples.

\begin{table}
    \centering
    \begin{tabular}{c | cc}
        \hline 
        \hline
        Quantity & Value(s) & \\
        \hline
        \hline
        $M_\ast$ & $M_\odot$ & \\
        $L_\ast$ & $1.2L_\odot$ & \\
        $M_0$ & $0.1M_\odot$ & \\
        \hline
        $R_0$ & [20, 40, 60, 80] au & \\
        $\alpha$ & $10^{-4},10^{-3}$ & \\
        $v_\mathrm{frag}$ & $\mathrm{3~ms^{-1},9~ms^{-1}}$ & \\
        \hline
        $t_0$ & [0.1, 0.25, 0.5, 0.75, 1] Myr & \\
        $a_\mathrm{p}(t_0)$ & [0.5, 1, 2, 5, 10, 20] au & \\
        \hline
    \end{tabular}
    \caption{Model parameters used in this study. Stellar parameters (top) are held constant, while disc (middle) and protoplanet (bottom) parameters are allowed to vary.}
    \label{tab:all_params}
\end{table}

\subsection{Central star and disc parameters}
We choose a Sun-like star with mass $M_{\ast}=1M_{\odot}$ and luminosity $L_{\ast}=1.2L_{\odot}$ and take the initial disk mass $M_{0}=0.1M_{\odot}$ and characteristic radii $R_{0}=[20, 40, 60, 80]\mathrm{~au}$. We also vary the disk parameters $\alpha=[10^{-4}, 10^{-3}]$ and $v_\mathrm{frag}=[3, 9]\mathrm{~m~s}^{-1}$ (see Table~\ref{tab:all_params}). These ranges correspond roughly to those in \citet{Krijt_Banzatti_2025}, and typical values for the turbulence are in line with \citet{rosotti2023}. For the disc's composition, the initial elemental and molecular compositions are identical to those in \citet[][Tables~1 and 2]{Schneider_Bitsch_2021} and are based on \citet{asplund2009}. For each combination of parameters, the initial surface density profile $\Sigma_\mathrm{g}(R)$, disc temperature profile\footnote{The temperature profile is calculated by considering both viscous heating and irradiation by the star, but \texttt{chemcomp} currently does not include temperatures evolving with time \citep[see][Appendix~B]{Schneider_Bitsch_2021}.} $T(R)$ (important for the snowline position), the pressure gradient $\eta(R)$ (important for pebble drift) and other relevant quantities are calculated self-consistently as detailed in \citet[][]{Schneider_Bitsch_2021}.

\begin{figure*}
    \centering
    \includegraphics[width=\linewidth]{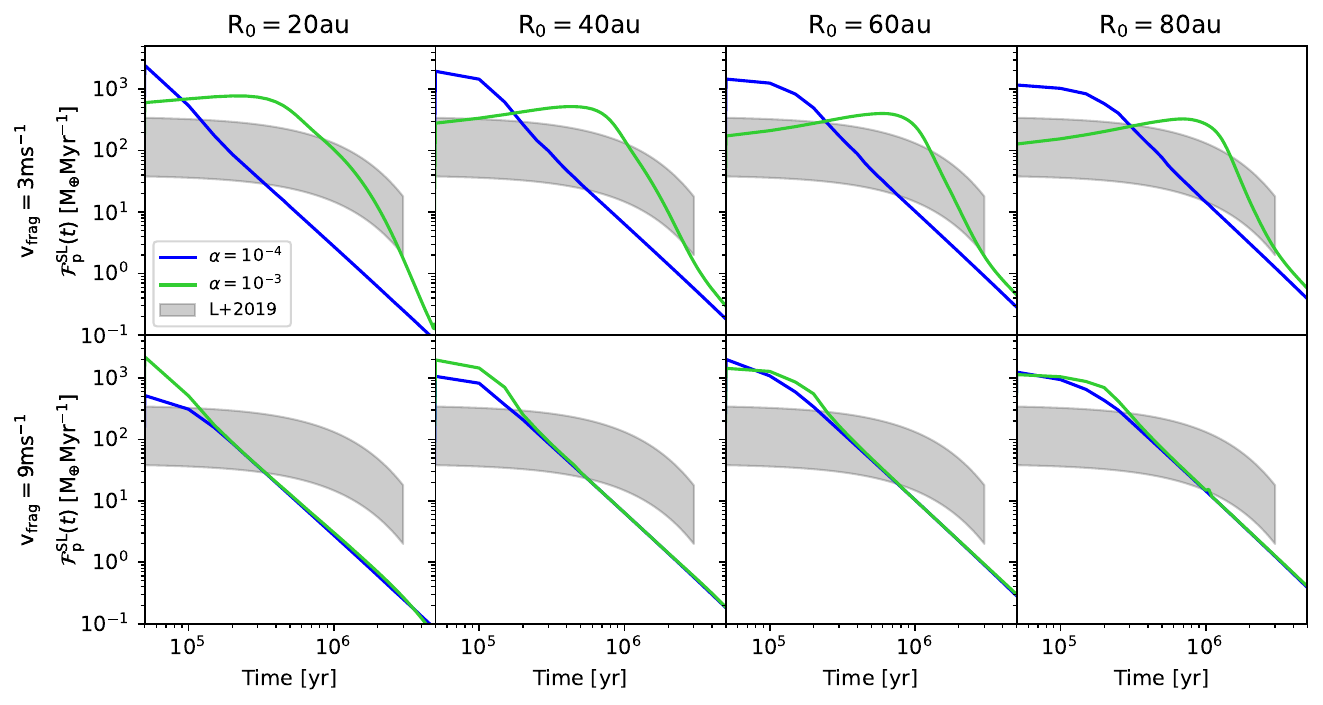}
    \caption{Snowline pebble mass fluxes over time for disks of varying $R_0$, $\alpha$ and $v_\mathrm{frag}$. The shaded region shows pebble mass fluxes assumed in the accretion simulations of \citet{lambrechts2019}.} 
    \label{fig:pebble_flux_vs_time_no_planets}
\end{figure*}

\subsection{The snowline pebble flux}
\label{sec:flux}
In general, dust evolution and pebble drift models (in the absence of complicating factors due to planets or substructure) predict radial pebble fluxes in the inner disc regions to first increase (as dust coagulation takes place), then plateau for some period (as the disc is close to steady-state), and then decrease gradually as the outer disc is drained of solids through drift \citep[see e.g.,][]{birnstiel2012,drazkowska_pp7}. We are primarily interested in the time-evolution of the mass flux near the (midplane) water snowline $\mathcal{F}^\mathrm{SL}_\mathrm{p}(t)$, i.e., at the radial position where $T(R)=150\mathrm{~K}$, which for our systems happens between ${\approx1{-}5\mathrm{~au}}$.

The pebble flux at any given location can be written as $\dot{m}_p=2\pi r v_r\Sigma_\mathrm{dust}$ \citep[][]{drazkowska2021} where $\Sigma_\mathrm{dust}$ is the surface density of solids and $v_r$ the pebble radial drift velocity. In \texttt{chemcomp}'s framework, dust coagulation/fragmentation is treated using an approach similar to the two-population model of \citet{birnstiel2012}, but due to the presence of multiple dust and molecular species the radial pebble flux is made up of multiple refractory (i.e., rocky) and volatile (i.e., icy) compounds whose surface densities are added together \citep[][Sect.~2.1, 2.2]{Schneider_Bitsch_2021}. We note here that the pebble flux very close to the water (or any other) snowline can become elevated by the high abundance of ice resulting from the re-condensation of outwardly-diffused vapour. While this effect is interesting from a planetesimal formation perspective \citep[see e.g.,][]{schoonenberg2017, drazkowska2017, andama2024, xenos2025}, we are mainly interested in how much matter arrives from the outer parts and hence opt to measure the pebble flux slightly further out (out at a position where $T(R)=100\mathrm{~K}$), where the re-condensation will have a minimal effect on the pebble surface density.

Figure~\ref{fig:pebble_flux_vs_time_no_planets} illustrates the diversity in snowline pebble fluxes seen across our model grid. In each panel, we see how the pebble mass flux across the snowline evolves throughout a disc's lifetime for different disk conditions in our model. We observe typical ranges of pebble flux, between $10^{-7} - 10^{-3} ~ \mathrm{M_\oplus / yr}$, which are similar to those reported in \citet{Romero-Mirza_2024,Krijt_Banzatti_2025} as discussed in more detail in Sects. \ref{sec:1Myr_fluxes} and \ref{sec:discussion}. It can be seen how certain aspects, such as the peak of the flux, can be delayed in larger disks. These are similar to results found in \citet[][Fig.~7]{drazkowska2021} and \citet[][Fig.~2]{Williams_Krijt_2025}, particularly how a larger disk, lower $v_\mathrm{frag}$ and higher $\alpha$ extends the time pebble flux remains at a higher value \citep[see also][]{Bitsch_Mah_2023}. The maximum pebble size in the case of fragmentation-limited growth scales as $s_\mathrm{max} \propto v_\mathrm{f}^2 / \alpha$ \citep{birnstiel2012,birnstiel2024}, so either a higher fragmentation velocity or weaker turbulence will lead to larger pebbles, which in turn result in faster drift and higher pebble fluxes early on ($t\sim0.1{-}1\mathrm{~Myr}$), which then in turn results in faster dust depletion and lower pebble fluxes at $t \gtrsim \mathrm{1Myr}$.

\subsection{Protoplanet injection}
For a given disc model, we insert (proto)planets (one planet per disc) at different times $t_0 = [0.1, 0.25, 0.5, 0.75, 1]\mathrm{~Myr}$ and a range of initial semi-major axes $a_\mathrm{p}(t_0) = [0.5, 1, 2, 5, 10, 20]\mathrm{~au}$. Their initial masses are calculated by \verb|chemcomp| to be the pebble transition mass, where pebble accretion starts to become efficient, given by
\begin{equation}\label{eq:pebble_transition_mass}
    M_\mathrm{t} = \sqrt{\frac{1}{3}}\frac{\Delta v^3}{G\Omega}
\end{equation}
where $\Delta v$ is the deviation of the gas azimuthal velocity from the Keplerian due to the radial pressure gradient in the disc, $\Omega$ is the Keplerian angular frequency, and $G$ is the gravitational constant \citep[see][Eqs.~3 and 29]{Schneider_Bitsch_2021}. This choice results in initial protoplanet masses increasing with their initial position $a_p$, with maximum initial masses of up to ${\approx}0.1~M_\oplus$ for those furthest out. For each planet, variations in disc parameters and the position and time at which the planet is inserted give rise to a diverse range of planet accretion and migration scenarios and hence planetary outcomes (e.g., final planet masses and orbital radius) \citep[e.g.,][]{Schneider_Bitsch_2021,savvidou_bitsch_2023}.

\subsection{Planet accretion and migration}\label{sec:accretion_migration}
Once injected, protoplanets grow by accreting drifting pebbles and/or gas, and can migrate radially as they interact with the ambient disc. For pebble accretion, \texttt{chemcomp} convolves the evolving pebble flux at the planet's location with a pebble accretion efficiency which depends primarily on the pebble Stokes number, the turbulence, and the protoplanet mass \citep[see][Sect.~2.6.1 for a detailed description]{Schneider_Bitsch_2021}. Both 2D and 3D regimes \citep[see e.g.,][]{johansen2017, ormel2018} are considered. 

A key mass scale in this story is the so-called pebble isolation mass, above which planets significantly perturb the local gas surface density, preventing pebbles from reaching their location and bringing a halt to pebble accretion. Based on \citet{bitsch2018}, \texttt{chemcomp} takes the pebble isolation mass to equal
\begin{equation}\label{eq:M_iso}
    M_\mathrm{iso} = 25 M_\oplus\times\left(\frac{H/r}{0.05} \right)^3 \left[ 0.66 + 0.34 \left(\frac{\log(10^{-3})}{\log(\alpha)} \right)^4\right],
\end{equation}
where $H$ is the gas scale-height, $r$ distance to the star, and $\alpha$ the turbulence. As $H/r$ typically increases with radius, the pebble isolation mass increases towards the outer disk \citep[see also][]{savvidou_bitsch_2023}.

For gas accretion, the accretion rate onto the planet is based on results from \citet{ikoma2000, machida2010}, and is capped by the rate at which the disc can viscously supply gas to the planet's horseshoe region \citep{ndugu2021}. For a complete description of the procedure and the assumptions therein we refer the reader to \citet[][Sect.~2.6.2]{Schneider_Bitsch_2021}.

Type I and type II planet migration are included as described in \citet[][Sect.~2.5]{Schneider_Bitsch_2021}, with rates that depend on planet mass and directions that can be inwards or outwards \citep[see also][]{ndugu2021}. We return to planet migration (and in particular the way in which the rates depend on viscosity) in more detail in Sects.~\ref{sec:t0_fluxes}.

\subsection{Feedback on the gas and pebbles}\label{sec:feedback}
Once planets exceed the pebble isolation mass, their presence will perturb the local gas density and lead to the creation of a gap and (if the gap is deep enough) an associated pressure maximum exterior to the gap. Since the pebbles' radial drift velocity scales with the pressure gradient, pebbles can become stuck at this pressure maximum, reducing the pebble mass flux reaching the disc regions interior to the planet's orbit. In \texttt{chemcomp}, the planet-induced gap in the gas surface density is achieved by introducing a planet-mass-dependent modulation in the local viscosity parameter $\alpha$. This modulation, the strength of which increases as the planetary mass increases, results in a reduced gas surface density in the gap while keeping the radial gas accretion rate constant \citep[see][Sect.~2.7]{Schneider_Bitsch_2021}.

\begin{figure*}
    \centering
    \includegraphics[width=\linewidth]{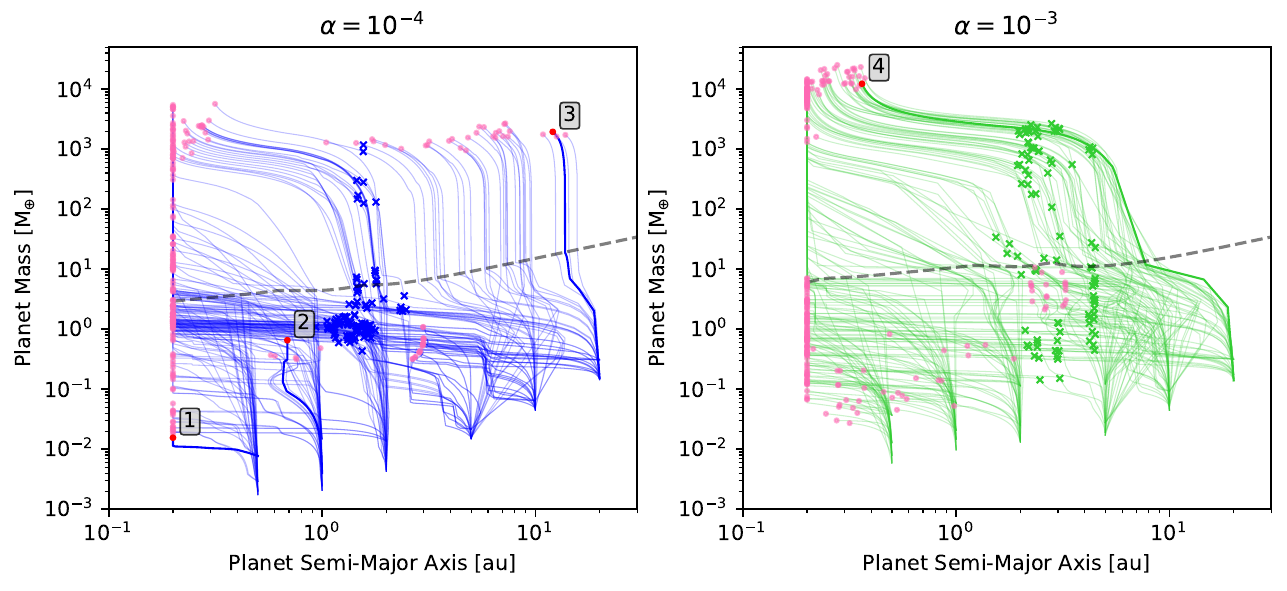}
    \caption{Planet growth tracks, mass vs semi-major axis. Pink dots represent the final planet mass and positions, with the numbered red dots showing selected planets numbered 1-4. Crosses show where the planet crosses the water snowline, if it does so. The black dashed line represents the approximate pebble isolation mass (Eq.~\ref{eq:M_iso}) for a disk with $R_0=40\mathrm{~au}$. The difference to other disks is minor, within a few $M_\oplus$.}
    \label{fig:planet_growth_tracks}
\end{figure*}

\subsection{Representative planet growth tracks}\label{sec:typical_planets}

We present in Fig.~\ref{fig:planet_growth_tracks} the planet growth tracks of 480 planets generated using the methodology outlined above, and illustrate the diversity in outcomes by discussing briefly the behaviour of four `typical' planets highlighted in the Figure and with initial and final properties summarized in Table~\ref{tab:4planets}.

\begin{table*}
    \centering
    \begin{tabular}{c | c c c c c | c c}
        \hline \hline
        Planet & $t_0$ [Myr] & $R_0$ [au] & $\alpha$ & $v_\mathrm{frag}$ [$\mathrm{m~s^{-1}}$] & $a_\mathrm{p}(t_0)~\mathrm{[au]}$ & $M_\mathrm{p}^\mathrm{final}$ [M$_\oplus$] & $a_\mathrm{p}^\mathrm{final}~\mathrm{[au]}$\\
        \hline
        1 & 1.0 & 20 & $10^{-4}$ & 9 & 0.5 & 0.0156 & 0.2\\
        2 & 0.5 & 40 & $10^{-4}$ & 3 & 1.0 & 0.657 & 0.69\\
        3 & 0.1 & 60 & $10^{-4}$ & 3 & 20 & 1950 & 12\\
        4 & 0.1 & 80 & $10^{-3}$ & 9 & 20 & 12300 & 0.36\\
        \hline
    \end{tabular}
    \caption{Initial conditions and final masses and locations for highlighted planets 1-4 discussed in Sect.~\ref{sec:typical_planets} and shown in Figures \ref{fig:planet_growth_tracks}, \ref{fig:M_v_flux_t0}, \ref{fig:M_v_flux_1Myr}.}
    \label{tab:4planets}
\end{table*}

Planet 1 is injected relatively late ($t_0=1\mathrm{Myr}$) and close-in in a small disc with low turbulence and high fragmentation velocity (see Table~\ref{tab:4planets}). For these disc parameters, the pebble flux by 1Myr has already significantly decreased (see Fig.~\ref{fig:pebble_flux_vs_time_no_planets}), and as a result Planet 1's growth is slow, and it ends up with a final mass well below $0.1M_\oplus$ as one of the smallest planets in our simulations. Planet 2 is inserted a bit earlier ($t_0=0.5\mathrm{Myr}$) just inside the snowline in an intermediate-sized disc with a lower fragmentation velocity. Due to the higher pebble flux around $t_0$ this planet can grow more easily, reaching $\sim0.6\mathrm{M_\oplus}$ before the simulation is stopped. Conversely, planet 3 is inserted very early (0.1~Myr) and far out, at $a_\mathrm{p}(t_0)=20\mathrm{~au}$ in a low-turbulence disc. Growing rapidly due to the large pebble flux combined with low $\alpha$, it crosses the pebble isolation mass and grows into a gas giant without migrating significantly inward, ending up at a semi-major axis of 12 au. Lastly, planet 4 highlights the role of turbulence. Inserted at a similar time and position as Planet 3, this planet also captures a lot of pebbles and reaches the pebble isolation mass in the outer disc, but due to the higher $\alpha$ Planet 4 migrated through the water snowline (see below) and ultimately ends up inside the inner 0.5 au where it accretes more gas until reaching a final mass exceeding $10^4~M_\oplus$. These four planets then highlight variations in outcomes that have been discussed previously in the context of pebble and gas accretion models (with or without migration) \citep[see e.g.,][]{bitsch2015,Schneider_Bitsch_2021,schneider2021b,drazkowska2021,savvidou_bitsch_2023, johnston2025}.

We also highlight in Fig.~\ref{fig:planet_growth_tracks} the position (and current mass) of a planet when it migrates through the water snowline (coloured crosses). Snowline positions can be seen to range from 1 to 5 au, roughly in line with snowline position estimates for disc models \citep[e.g.,][]{mulders2015,savvidou2020,kim2025}. In our models, the spread in snowline positions is caused by the variation in the viscous heating term (which scales with $\alpha$ and opacity \citep[see][Appendix~B]{Schneider_Bitsch_2021} which cause variations in the disk temperature even if the stellar mass and luminosity is kept constant. The locations of the x's in Fig. \ref{fig:planet_growth_tracks} will be relevant for Sect.~\ref{sec:fluxes_with_planets}, where we discuss the planets' impact on the snowline pebble flux, as only planets that are (a) massive, and (b) (well) outside the snowline are expected to influence the pebble flux arriving at the inner disc.

\section{Results}
We now turn to discussing the connection between the final planet properties (e.g., final mass and position as indicated by the pink points in Fig.~\ref{fig:planet_growth_tracks}) and snowline pebble flux (as depicted in Fig.~\ref{fig:pebble_flux_vs_time_no_planets}). We will follow two approaches that tell different parts of the story. First, we compare planet outcomes to the snowline pebble flux \emph{measured at the time of protoplanet insertion} (Sect.~\ref{sec:t0_fluxes}). This approach highlights primarily the impact of the snowline pebble flux (and other disc properties) on the planet's ability to grow. Additionally, in Sect.~\ref{sec:1Myr_fluxes}, we compare planet outcomes to the snowline pebble flux \emph{as evaluated around $1~\mathrm{Myr}$} which allows for a more direct connection of observations as this time is closer to the ages sampled in e.g., \citet{Krijt_Banzatti_2025}. We discuss the impact of planets (if already sufficiently massive by 1~Myr) on the snowline pebble flux in Sect.~\ref{sec:fluxes_with_planets}.

\subsection{Pebble flux at the time of (proto)planet injection}\label{sec:t0_fluxes}

\begin{figure*}
    \centering
    \includegraphics[width=\linewidth]{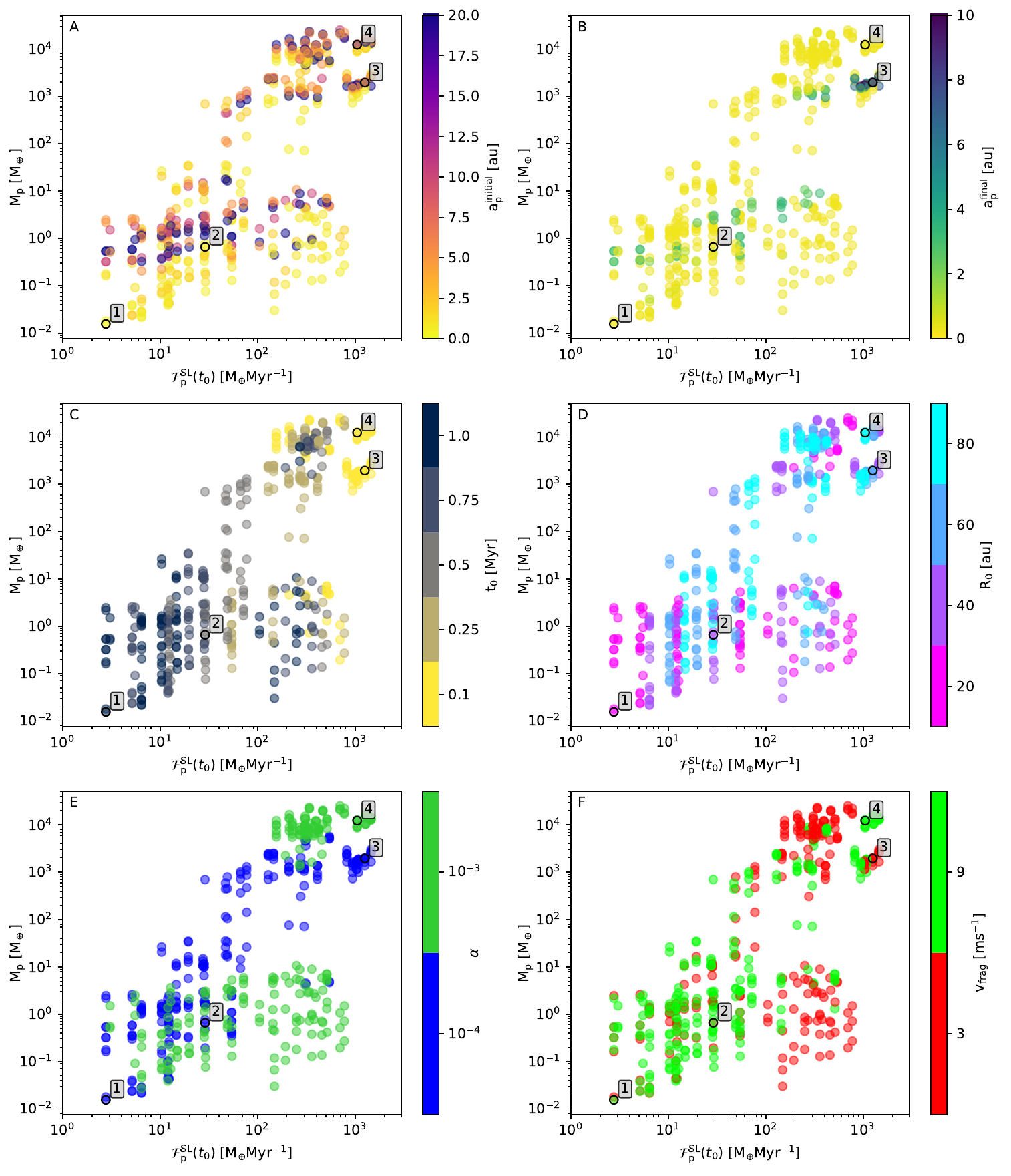}
    \caption{Final planet mass vs pebble flux across the snowline at time of insertion. Plots A and B are coloured by the planet's initial and final semi-major axes respectively. Plot C is coloured by the planet's insertion time $t_0$, and plot D is coloured by the characteristic disk radius $R_0$. Plot E is coloured by the disk's $\alpha$ parameter, and plot F is coloured by the fragmentation velocity $v_\mathrm{frag}$. All plots contain outlined numbered points, which correspond to the points numbered in Fig.~\ref{fig:planet_growth_tracks}.}
    \label{fig:M_v_flux_t0}
\end{figure*}

Figure~\ref{fig:M_v_flux_t0} shows final planet mass vs. the snowline pebble mass flux (evaluated at the time of protoplanet injection $t_0$) as a function of a range of disc and protoplanet parameters (various colour bars). We also highlight in each panel the positions of the four planets discussed in Sect.~\ref{sec:typical_planets} to aid in the interpretation.

First and foremost, just from the distribution of the points in any of the panels, we see there is a clear correlation between the (maximum) final planet mass and snowline pebble flux: protoplanets inserted when the pebble flux is below ${\approx}50{-100}M_\oplus\mathrm{~Myr^{-1}}$ can not reach masses exceeding a few $10M_\oplus$. This can be explained by planets in discs with a lower pebble flux struggling to grow quickly early on, and thus not being able to capture significant gas envelopes during the gas disc lifetime.

On top of this general trend however there is considerable scatter. Indeed, even in discs with pebble fluxes approaching ${\approx}10^3~M_\oplus\mathrm{~Myr^{-1}}$ there are planets that end up with masses of (just) $0.1M_\oplus$. There can be various reasons for this. First, it is worth stressing that while the pebble flux is measured \emph{at the water snowline} (as we want to compare fluxes inferred from JWST observations), protoplanets are inserted with initial positions ranging from 0.5 and 20 au (see Sect.~\ref{sec:typical_planets} and Fig.~\ref{fig:M_v_flux_t0}A). As such, there may be differences between the pebble flux at the planet position (most relevant for its growth) and the one at the snowline, although we expect these differences to be minimal in the inner $\sim10\mathrm{au}$ for most of the disc's lifetime \citep[see e.g.][Fig. 1]{mulders_2021}. Second, and likely more important, is that the rate of pebble accretion depends not solely on the pebble flux but also on the pebble accretion efficiency, which itself is a complex function of also the pebble Stokes numbers, the turbulence strength $\alpha$, the protoplanet mass (relative to the star) and the disc aspect ratio \citep[e.g.,][]{johansen2017,ormel2018} -- all factors that vary from disc to disc and planet to planet. We discuss some of these factors in more detail here.

From Fig.~\ref{fig:M_v_flux_t0}C, we find that planet injection time plays a critical role in the formation of gas giants with $M_\mathrm{p} \gtrsim 10^2 M_\oplus$: planets injected early have a higher chance of experiencing high pebble fluxes (see also Fig.~\ref{fig:pebble_flux_vs_time_no_planets}) and thus a higher probability of growing into gas giants. Similar results, i.e., earlier insertion times resulting in larger planets, were found by \citet[][Fig.~3]{drazkowska2021} (for static planets) and \citet[][Fig.~5]{savvidou_bitsch_2023} (for migrating planets).

Most planets migrate inwards by considerable distances, owing to rapid type II migration -- especially in the $\alpha=10^{-3}$ case (see also Fig.~\ref{fig:planet_growth_tracks}). Those that start further out gain the most mass as they travel through the disk. This is seen clearly Fig.~\ref{fig:M_v_flux_t0}B, which shows the final positions of the majority of the planets to be ${\lesssim}1\mathrm{~au}$. Additionally, the majority of the highest-mass planets grew from protoplanets that were inserted early (top-right region of panel C).

In plot D we find that the planet mass doesn't necessarily correspond to the disk size, however the lowest pebble fluxes are in the smallest disks, and hence produce the smallest planets. However, we also see that these smaller disks can produce the largest planets, with more material in the vicinity of the planet resulting in higher accretion rates, which was also seen in \citet[][]{savvidou_bitsch_2023}.

Looking at the effect of changing the turbulence (panel E), we notice that there is positive correlation between final planet mass and pebble flux across the entire parameter space for the low turbulence case ($\alpha=10^{-4}$), while the $\alpha=10^{-3}$ simulations result in two populations: giant planets (found only at high pebble fluxes) and $\lesssim10M_{\oplus}$ planets (found across all pebble fluxes). Similar behaviour is observed in \citet[][Fig.~5]{savvidou_bitsch_2023}, where it is explained through a combination of faster gas accretion (as gas accretion is limited by the accretion rate through the disc, which itself scales with $\alpha$) and faster (type II) planetary migration for higher viscosities, resulting in fewer planets being left behind at intermediate masses \citep[][ and Sect.~4.4]{savvidou_bitsch_2023}. The effects of faster growth and migration can also be seen in Fig. \ref{fig:planet_growth_tracks}, which shows giants planets outside of 1 au only in the $\alpha=10^{-4}$ case.

Similarly, plot F shows a clearer correlation between final planet mass and snowline pebble flux for the $v_\mathrm{frag}=9\mathrm{~m~s^{-1}}$ case, while for $v_\mathrm{frag}=3\mathrm{~m~s^{-1}}$ an additional population of low-mass planets exists even for high snowline pebble fluxes ${>}100~M_\oplus/\mathrm{Myr}$. This population of protoplanets, which struggles to grow despite high pebble mass fluxes being present (at least during the time of insertion) may be explained by low pebble accretion \emph{efficiencies} as the high turbulence levels (panel E) and low fragmentation velocity (and thus lower Stokes numbers in the fragmentation regime) (panel F) both contribute to decreased pebble accretion efficiencies \citep[e.g.,][Fig.~4]{ormel2018}. The late injection times for many of these planets (panel C) also play a role, as these imply the pebble fluxes -- while high at $t_0$ may start to decrease relatively quickly (see Fig.~\ref{fig:pebble_flux_vs_time_no_planets}).

\subsection{Pebble flux at 1 Myr}\label{sec:1Myr_fluxes}
\begin{figure*}
    \centering
    \includegraphics[width=\linewidth]{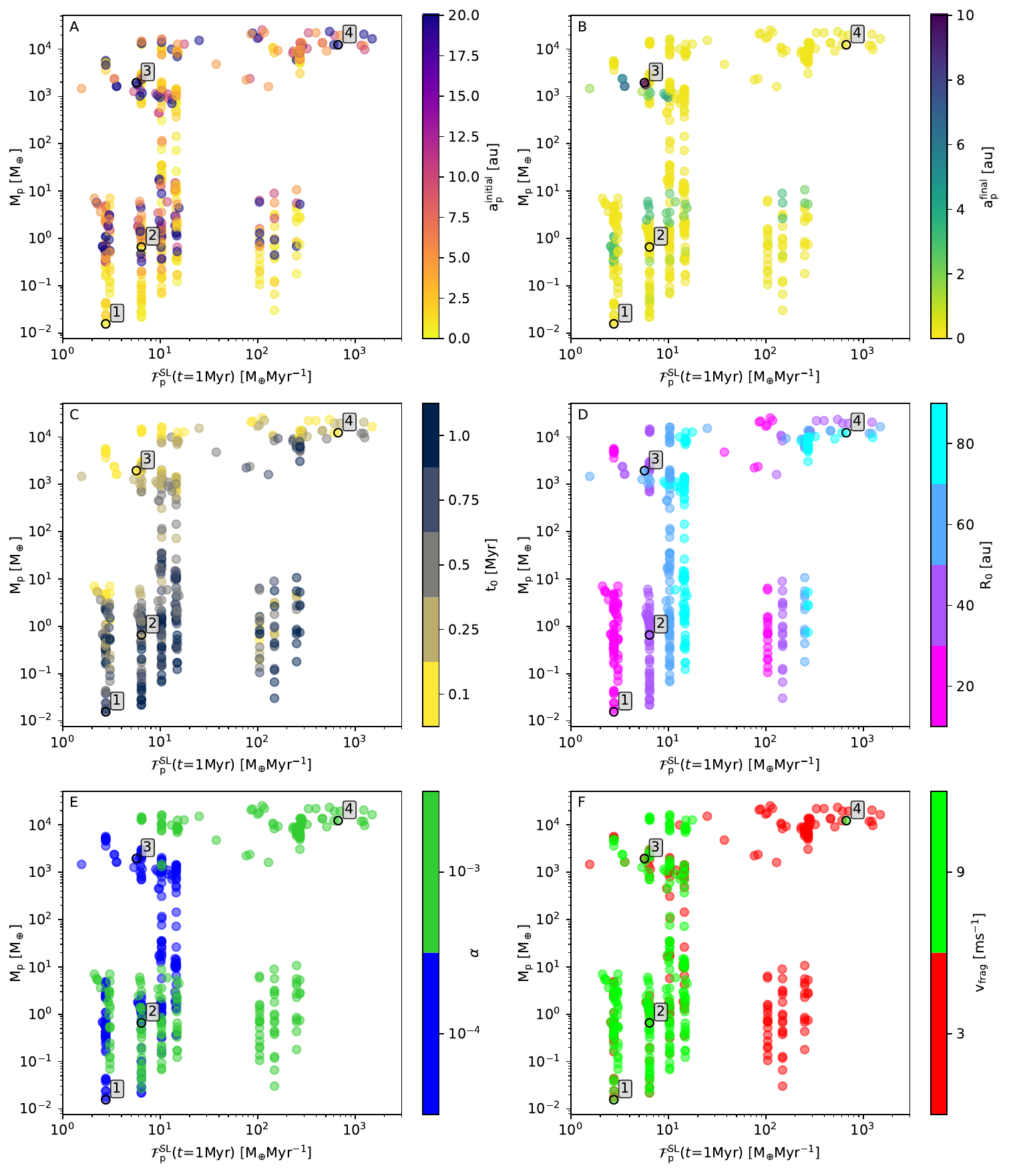}
    \caption{Final planet mass vs pebble flux across the snowline at 1 Myr. Plots A and B are coloured by the planet's final and initial semi-major axes respectively. Plot C is coloured by the planet's insertion time $t_0$, and plot D is coloured by the characteristic disk radius $R_0$. Plot E is coloured by the disk's $\alpha$ parameter, and plot F is coloured by the fragmentation velocity $v_\mathrm{frag}$. All plots contain outlined numbered points, which correspond to the points numbered in Fig.~\ref{fig:planet_growth_tracks}.}
    \label{fig:M_v_flux_1Myr}
\end{figure*}

The pebble mass fluxes inferred from the excess cold water emission in JWST/MIRI spectra \citep[e.g.,][]{Romero-Mirza_2024} so far primarily sample systems with ages between ${\approx}~0.5-1.5\mathrm{~Myr}$ \citep[][Table~1]{Krijt_Banzatti_2025}, and it remains unclear how far planet formation (especially in the inner disc) has proceeded in these systems. To investigate the predictive power of these observations, we show in Fig.~\ref{fig:M_v_flux_1Myr} the same synthetic planet population as in Fig.~\ref{fig:M_v_flux_t0} but this time using the snowline pebble flux evaluated at 1 Myr. With planet injection times between 0.1 and 1 Myr, some of these systems will already contain gas giants by the time the pebble flux is measured\footnote{The complete planet mass and orbital period distribution as present at $t=1~\mathrm{Myr}$ is provided in Appendix~\ref{sec:planet_pop_at_1Myr}.}, in which case the snowline pebble flux can be affected dramatically by the impact of the giant planet on the gas surface density profile and pebble drift behaviour. This `feedback' of the growing planet on the pebble flux through the disc, included self-consistently in \texttt{chemcomp}, can be quite dramatic and is discussed in the next Sect.~\ref{sec:fluxes_with_planets}.

Comparing Figs.~\ref{fig:M_v_flux_1Myr} and \ref{fig:M_v_flux_t0} it is clear that the correlation between final planet mass and pebble flux at 1 Myr is significantly weaker and obfuscated by the variations in the various disc and planet parameters. Nonetheless, we can point out a few interesting observations.

When looking at the final planet mass against the pebble flux measured at 1 Myr in Fig. \ref{fig:M_v_flux_1Myr} we can relate the observations of disks to our simulations, and see how planets may affect the disk. Here we find several vertical lines along discrete values of the pebble flux. In plot C we can see that these lines are mostly produced by later insertion times, suggesting that these planets haven't had a chance to affect the disk's pebble flux. We can see the largest planets are generally inserted earlier ($\leq0.5\mathrm{Myr}$), and the pebble fluxes are more spread out, suggesting these planets could be affecting the pebble flux as seen in Fig. \ref{fig:pebble_flux_vs_time_with_planets}.

In plots E and F we can see a split between pebble flux for different values of $\alpha$ and $v_\mathrm{frag}$, which once again comes from the late insertion time of the planets, and consequently shows a mostly unperturbed flux as would be expected from Fig. \ref{fig:pebble_flux_vs_time_no_planets}. Focusing on only the largest planets we can again find that they can come from any disk size, and generally start far out and migrate inwards, as seen in plots A and B. A small correlation may be seen between the pebble flux at 1 Myr and the final planet mass for the largest planets. But, overall there isn't as clear a correlation between the planet mass and pebble flux at 1 Myr as there was when taking the pebble flux at the insertion time in Fig. \ref{fig:M_v_flux_t0}. 

\subsection{Pebble flux evolution in planet-hosting discs}\label{sec:fluxes_with_planets}
\begin{figure*}
    \centering
    \includegraphics[width=\linewidth]{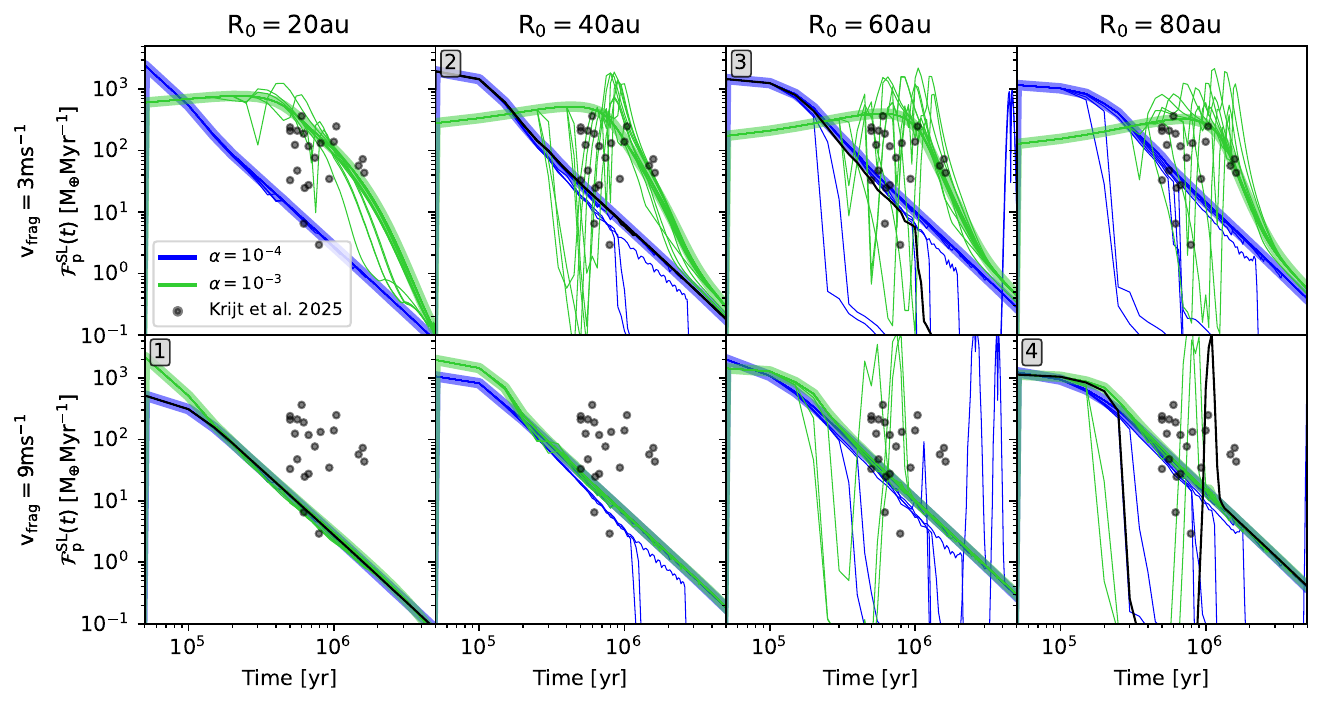}
    \caption{Pebble fluxes over time for disks of varying characteristic radius $R_0$ (left to right), turbulence level $\alpha$ (blue vs. green) and fragmentation velocity $v_\mathrm{frag}$ (top and bottom). The thicker lines show the flux from Fig. \ref{fig:pebble_flux_vs_time_no_planets} without planets (see Sect.~\ref{sec:flux}) and the thin lines show the fluxes in the same discs but now co-evolving with growing individual planets. The numbered black lines are for the typical planets discussed in Sect. \ref{sec:typical_planets}. In the simulations where significantly massive planets form, the pebble flux profiles display a sharp decrease in the pebble flux, sometimes followed by a single spike or surge (especially in the $\alpha=10^{-3}$ case), the timing of which varies from simulation to simulation (see Sect.~\ref{sec:fluxes_with_planets}).
    We also overlay the pebble mass fluxes inferred from cold water excess glimpsed by JWST/MIRI as described in \citet[][]{Krijt_Banzatti_2025}.}
    \label{fig:pebble_flux_vs_time_with_planets}
\end{figure*}

Giant planets are expected to carve out gaps in the gas surface density that can -- locally -- act as pebble traps/filters with various degrees of efficiency \citep[e.g.,][]{weber2018,bitsch2018,stammler_2023, huang_2025, savvidou_bitsch_2025}, affecting also the flux of pebbles arriving in the inner disc \citep[e.g.,][]{kalyaan_2021, kalyaan_2023, mah_2024, pinilla_2024, easterwood_2024}. To quantify this feedback for our simulations we show in Fig.~\ref{fig:pebble_flux_vs_time_with_planets} once more snowline flux vs. time, in the same format as in Fig.~\ref{fig:pebble_flux_vs_time_no_planets} but this time extracting the mass flux from the simulations with planets for which the impact of the planets on the disc-wide pebble flux is calculated self-consistently as the planets grow and migrate. For context, we show also the 21 discs described in \citet{Krijt_Banzatti_2025}. Current dust disc radii for these systems are known and fall between ${\approx}30-300\mathrm{~au}$ \citep[][Fig.~1]{Krijt_Banzatti_2025}, however, these radii -- as shaped by substructure -- are expected to deviate from initial gas disc characteristic sizes and we therefore plot all discs in each panel in Fig.~\ref{fig:pebble_flux_vs_time_with_planets}.

We see that the formation of a planet may greatly decrease the pebble flux after a few $10^5\mathrm{~yr}$, by factors of up to and exceeding two orders of magnitude. This particularly evident in larger disks, where more mass is likely to be present exterior to the giant planet, an effect captured in the analytical model of \citet[][Eq.~2]{Krijt_Banzatti_2025}. Additionally, we see that in disks with lower $\alpha$ (blue lines) the pebble flux is usually decreased essentially permanently, whereas in discs with $\alpha=10^{-3}$ the flux is often decreased only for a period of ${\approx}10^{5-6}\mathrm{~yr}$ before climbing again and even spiking briefly before returning to typical (non-planet) value. We attribute this behaviour to massive planets migrating inward and approaching and passing through the midplane snowline. Once the planets cross the snowline, the pebbles trapped in the pressure bump exterior to the planet will also follow, leading to a temporary spike in the pebble mass flux \citep[see also][]{Eberlein_2024}. Once this spike has passed, the planets -- now well inside the snowline -- have a negligible impact on the the snowline pebble flux. Fig.~\ref{fig:planet_growth_tracks} already showed that many of our planets cross the snowline, but what is key is that: (1) giant planets in the $\alpha=10^{-4}$ models have a higher probability of staying in the outer disc, and (2) \emph{if} crossing the snowline, the planet masses in the $\alpha=10^{-4}$ models are generally below the pebble isolation mass (shown as dashed lines in Fig.~\ref{fig:planet_growth_tracks}). These two effects together explain why pebble flux decreases in our simulations are more short-lived particularly in the $\alpha=10^{-3}$ models, and why this period usually ends with a brief `spike' in pebble flux.

\section{Discussion}\label{sec:discussion}
We now turn to discussing how the above results can connect to recently derived snowline pebble fluxes in a handful of $\sim$1Myr-old protoplanetary disks, and the limitations within observations and our models.

\subsection{Connecting planet formation to JWST/MIRI-derived snowline pebble fluxes}\label{sec:link_to_JWST}
One of the main drivers behind this study was provided by the snowline pebble mass flux constraints derived from JWST/MIRI measurements of cold water excess as presented in \citet{Romero-Mirza_2024} and \citet{Krijt_Banzatti_2025}.

For comparison, we show in Fig.~\ref{fig:pebble_flux_vs_time_with_planets} the pebble fluxes and ages of the twenty-one discs studies studied in \citet{Krijt_Banzatti_2025}. It is important to note here, however, that the system ages and inferred pebble fluxes used for the comparison in Fig.~\ref{fig:pebble_flux_vs_time_with_planets} carry significant uncertainties: For the ages the typical uncertainty is ${\sim}0.5\mathrm{~Myr}$ \citep[discussed in][Sect.~2]{Krijt_Banzatti_2025}, while the pebble flux carries uncertainties associated with (1) estimating the mass of the surface cold water reservoir\footnote{Which involves either fitting slab models \citep{Romero-Mirza_2024} or using empirical relations and measurements of specific water line flux ratios \citep[][Appendix~A]{Krijt_Banzatti_2025}.} and (2) converting that mass to a midplane pebble flux. The latter in particular can be complicated as pebbles also deliver dust which may shift the observable layers further up \citep[see e.g.,][]{houge2025a, sellek2025}. Nonetheless, our simulations, which self-consistently solve for dust coagulation and transport in discs with a (single) growing and migrating planet, appear in qualitative agreement with the toy models presented in \citet[][Sect.~3.4]{Krijt_Banzatti_2025} in that the data seem to rule out the $v_\mathrm{frag}=9\mathrm{~m/s}$ scenario as being dominant as it fails to reproduce typical pebble fluxes around $t \approx \mathrm{1~Myr}$. Instead; a relatively high turbulence and low fragmentation velocity is favoured. This is in line with recent studies attempting to reproduce multi-wavelength ALMA continuum observations in outer disc regions, which rule out combinations of high $v_\mathrm{frag}$ and low $\alpha$ \citep[e.g.,][]{jiang2024,tong2025,Williams_Krijt_2025}.

The presence of (giant) planets outside the snowline does -- as expected \citep[e.g.,][]{bitsch2021} -- act to reduce the pebble flux at the snowline by multiple orders of magnitude (cf. Figs.~\ref{fig:pebble_flux_vs_time_with_planets} and \ref{fig:pebble_flux_vs_time_no_planets}), an effect most evident in our simulations for discs with $R_0 \gtrsim 40\mathrm{~au}$. Interestingly, we also find that the reduction in pebble flux is limited to a period of ${\lesssim}\mathrm{~1~Myr}$, as on longer timescales the giant planet will migrate through and inside the snowline (see also Fig.~\ref{fig:planet_growth_tracks}), a process culminating in a brief increase in the snowline pebble flux (particularly evident in the $\alpha=10^{-3}$ simulations in Fig.~\ref{fig:pebble_flux_vs_time_with_planets}, see also \citealt{Eberlein_2024}). We do not suggest these brief spikes are responsible for any of the observed values\footnote{In fact, the methodology for estimating pebble mass fluxes from cold water excess developed in \citet{Romero-Mirza_2024} is expected to break down when there is a gap-opening giant planet in the immediate vicinity.} but this behaviour serves as a reminder that massive planets may perturb the inner disc evolution in multiple ways. We note that the ability of giant planets to migrate all the way to the snowline may be reduced if planet-planet interactions are included (see Sect.~\ref{sec:limitations}).

So what can the measured pebble fluxes say about the outcomes of planet formation in the inner regions? As shown in Fig.~\ref{fig:M_v_flux_t0}, the pebble mass flux \emph{at the time of protoplanet formation} does correlate with accretion outcomes (in this case planet mass), although system-to-system variations in disc parameters introduce significant scatter\footnote{As noted before, some of the combinations presented in this work are less likely then others (e.g., see discussion about low turbulence and high fragmentation velocity earlier). Including new constraints in future studies may help to reduce the scatter seen in Fig.~\ref{fig:M_v_flux_t0} and \ref{fig:M_v_flux_1Myr}.}. For example, focussing on planets whose cores were injected around $t_0=1\mathrm{~Myr}$ in Fig.~\ref{fig:M_v_flux_t0}C, it seems only planets forming in discs with snowline pebble fluxes ${\gtrsim}50{-}100~M_\oplus/\mathrm{Myr}$ have the potential of growing into gas giants, although this outcome is not guaranteed and growth may stall below $M_\mathrm{p}\sim1{-10}M_\oplus$, especially in small discs, discs with higher turbulence, and discs with low $v_\mathrm{frag}$ (see bottom-right regions in Fig.~\ref{fig:M_v_flux_t0} panels D, E, and F). If -- in addition to snowline pebble flux -- global individual disc parameters can be reasonably estimated (e.g., disc size, or the turbulence), these findings suggest planet formation outcomes can to some degree be predicted. 

However, the uncertainty in the \emph{timing} of protoplanet formation has a large impact on such predictions, particularly since current pebble flux constraints are all for ${\approx}\mathrm{Myr}$-old discs (cf. Figs.~\ref{fig:M_v_flux_1Myr} and \ref{fig:M_v_flux_t0}). In order to be able to make more quantitative predictions for planetary outcomes based on snowline pebble flux measurements, an improved understanding of (a) the timing of protoplanet formation \citep[and its likely dependence on formation location and disc parameters, see e.g.,][]{voelkel2020,voelkel2021} as well as (b) snowline pebble mass flux measurements for systems significantly younger/older then those used in \citet{Krijt_Banzatti_2025} appear to be key.

A final interesting point concerns the duration during which growing planets can exist \emph{before} significantly impacting the snowline pebble flux, i.e., how long can planets stay `unnoticed' \citep[see also][]{Nazari2025}? In our simulations, the mass scale at which planets significantly perturb the gas disc is essentially the pebble isolation mass given by Eq.~\ref{eq:M_iso} and shown in Fig.~\ref{fig:planet_growth_tracks}. With pebble accretion (and migration) timescales depending on disc parameters and planet insertion times (as discussed in Sect.~\ref{sec:typical_planets}), the time it takes to reach $M_\mathrm{iso}$ varies, but some trends remain: Focussing exclusively on planets that start growing outside the water snowline, the average time to grow to $M_\mathrm{iso}$ after being inserted equals $\langle t_{M_\mathrm{iso}}-t_0\rangle \approx0.5~\mathrm{Myr}$ for the planets in $\alpha=10^{-3}$ discs, and almost twice as long ($\langle t_{M_\mathrm{iso}}-t_0\rangle \approx0.9~\mathrm{Myr}$) for planets growing in $\alpha=10^{-4}$ discs. Folding in the variations in planet insertion times $t_0$, this results in typical disc ages of $\langle t_{M_\mathrm{iso}}\rangle \approx0.8~\mathrm{Myr}$ ($\alpha=10^{-3}$), and $\langle t_{M_\mathrm{iso}}\rangle \approx1.3~\mathrm{Myr}$ ($\alpha=10^{-4}$). Another way of showing these differences is by focussing on $t=1~\mathrm{Myr}$, at which time we find that $76\%$ of planets in $\alpha=10^{-3}$ have already reached $M_\mathrm{iso}$, compared to only $45\%$ in the discs with lower turbulence. So while considerable variation exists from planet to planet, this shows that in discs with low turbulence levels, the slower growth can result in planets staying undetected for longer -- at least from a snowline pebble flux point of view. During those periods where planets are still growing towards $M_\mathrm{iso}$, the pebble flux will still play a key role in shaping planet accretion (e.g., Fig.~\ref{fig:M_v_flux_t0}), but the feedback of the planets on the pebble flux (Fig.~\ref{fig:pebble_flux_vs_time_with_planets}) is expected to be minimal.

\subsection{Model limitations and future extensions}\label{sec:limitations}

We also comment briefly on some of the limitations of our model as outlined in Sect.~\ref{sec:model}. First, our model is limited to one planet per disc, and thus cannot capture direct or indirect planet-planet interactions. An interesting (indirect) effect for example could be that planets growing interior to any present giant planets will have their pebble accretion rates decreased due to the reduced pebble flux. This process was recently modelled by \citet{mccloat2025}, who found that growing masses in the outer disk are sufficient to prevent inner masses reaching the pebble isolation mass. However, \citet{Eberlein_2024} found this to not be the case due to outer planets reaching the pebble isolation mass later, when the inner planets would have already been formed. This again brings in to question the timing of protoplanet formation, and the way that process itself is regulated by the drifting pebbles \citep[e.g.,][]{voelkel2020,voelkel2021}. The lack of planet-planet gravitational interactions could also be a contributor to the reason most of our giant planets end up very close to their star, as we only see outward migration for smaller planets by construction, due to the heating torque which causes outward migration for planets up to a few Earth masses \citep[e.g.][]{benitez-llambay_2015,Masset_2017}.

Additionally, we note that the treatment of dust and pebble coagulation and drift in the \texttt{chemcomp} simulations we employed is based on the two-population model of \citet{birnstiel2012} and a 1D (vertically-integrated) disc model \citep{Schneider_Bitsch_2021}. Recent efforts that take into account the full dust size distribution and/or are performed in 2D show that these approximations can lead to variations in e.g., gap leakiness \citep[e.g.,][]{drazkowska2019,stammler_2023,Pfeil2025}, but we do not expect the results to change qualitatively. 

Finally, we do not self-consistently model the formation of planetesimals or their subsequent growth into (proto)planets of masses up to the mass given by Eq.~\ref{eq:pebble_transition_mass}. Instead, we effectively treat the planet injection time and position as free parameters (see Sect.~\ref{sec:typical_planets}). Diverse global (i.e., disc-wide) models predict very different formation times and positions \citep[as reviewed in][Sect.~3.1.4]{drazkowska_pp7}, we expect these processes to depend on other disc parameters such as $\alpha$ and $v_\mathrm{frag}$ \citep{birnstiel2016,drazkowska2018}. Thus, as some of the parameter combinations explored in this study will be more or less likely then others, the synthetic planet population resulting from them should not be compared directly to e.g., to exoplanet occurrence rates.

\section{Conclusions}
We have used \texttt{chemcomp} \citep{Schneider_Bitsch_2021,schneider2021b} to model joint disc, dust, and pebble evolution (Fig.~\ref{fig:pebble_flux_vs_time_no_planets}) together with planet accretion and migration (Fig.~\ref{fig:planet_growth_tracks}) to investigate more quantitatively the link between the snowline pebble flux on one hand, and final planet properties (e.g., mass, orbital radius) on the other. Our main conclusions can be summarized as follows:
\begin{enumerate}
    \item The (maximum) final planet mass strongly correlates with the water snowline pebble flux at the time of protoplanet insertion, while significant scatter is introduced when variations in disc parameters (e.g., $\alpha$, $v_\mathrm{frag}$, $R_0$) and/or different protoplanet insertion locations are considered (Fig.~\ref{fig:M_v_flux_t0}).
    \item For turbulence levels of $\alpha=10^{-4}$, the final planet mass vs. pebble flux correlation is essentially continuous, while in discs with $\alpha=10^{-3}$ the correlation behaves almost like a step function, with snowline pebble mass fluxes exceeding $100~M_\oplus/\mathrm{Myr}$ (at the time of protoplanet insertion) being necessary for the formation of gas giants (Fig.~\ref{fig:M_v_flux_t0}E).
    \item When comparing instead final planet masses to snowline pebble fluxes specifically around 1 Myr (for which most of the current constraints exist, see e.g., \citealt{Krijt_Banzatti_2025}), the variation in planet insertion times and various disc parameters allows for a broad range of outcomes and the predictive power of the snowline pebble flux is diminished (Fig.~\ref{fig:M_v_flux_1Myr}). Some of this degeneracy may be alleviated if for example additional constraints on disc turbulence level, disc size, present sub-structure, and/or typical protoplanet formation times can be obtained.
    \item Once grown to the pebble isolation mass, the planets evolving in our simulations can significantly perturb the snowline pebble flux, especially if they are located far out (i.e., well outside the snowline) and form early enough to be able to block in their wakes a significant mass of pebbles (cf. Fig.~\ref{fig:pebble_flux_vs_time_with_planets} and Fig.~\ref{fig:pebble_flux_vs_time_no_planets}).
    \item Intriguingly, our one-planet-per-disc simulations for $\alpha=10^{-3}$ result in the reduction of snowline pebble flux being only temporary, with giant planet growing and migrating quickly, and being unable to affect the snowline pebble flux once they have migrated to the innermost 1 au (Figs.~\ref{fig:pebble_flux_vs_time_with_planets} and \ref{fig:planet_growth_tracks} and Sect.~\ref{sec:link_to_JWST}).
\end{enumerate}
Generally speaking, the snowline pebble fluxes in our planet-containing simulations broadly overlap with those inferred from JWST/MIRI cold water excess studies \citep[Fig.~\ref{fig:pebble_flux_vs_time_with_planets}, and][]{Romero-Mirza_2024, Krijt_Banzatti_2025}, but we refrain here folding in additional constraints for all those discs (e.g., disc sizes, details of substructure, etc.), focussing instead on the impact these pebble fluxes have on (potentially) present protoplanets (Figs.~\ref{fig:M_v_flux_t0} and \ref{fig:M_v_flux_1Myr}) while Fig.~\ref{fig:pebble_flux_vs_time_with_planets} highlights how valuable additional pebble flux constraints -- especially for younger (${\lesssim}0.5~\mathrm{Myr}$) and older (${\gtrsim}2~\mathrm{Myr}$) disks could be for discriminating between models. To conclude, our simulations show that the currently observed pebble fluxes can be used to constrain planet formation simulations, and emphasize that diverse planet formation via pebble accretion is in agreement with the currently available constraints from JWST.

\section*{Acknowledgements}
The authors would like to thank Joe Williams for helpful suggestions and comments. DA would like to thank Scarlett Hutton, Caleb Ostler, and Alice Smith for their support and contributions during the MPhys project at the University of Exeter, from which this paper developed. The following Python packages were used for this study: \texttt{matplotlib} \citep[][]{matplotlib}, \texttt{numpy} \citep[][]{numpy}.

\section*{Data Availability}

The data underlying this article will be shared on reasonable request to the corresponding author.



\bibliographystyle{mnras}
\bibliography{example} 




\appendix

\section{Planet population at 1 Myr}\label{sec:planet_pop_at_1Myr}
\begin{figure*}
    \centering
    \includegraphics[width=\linewidth]{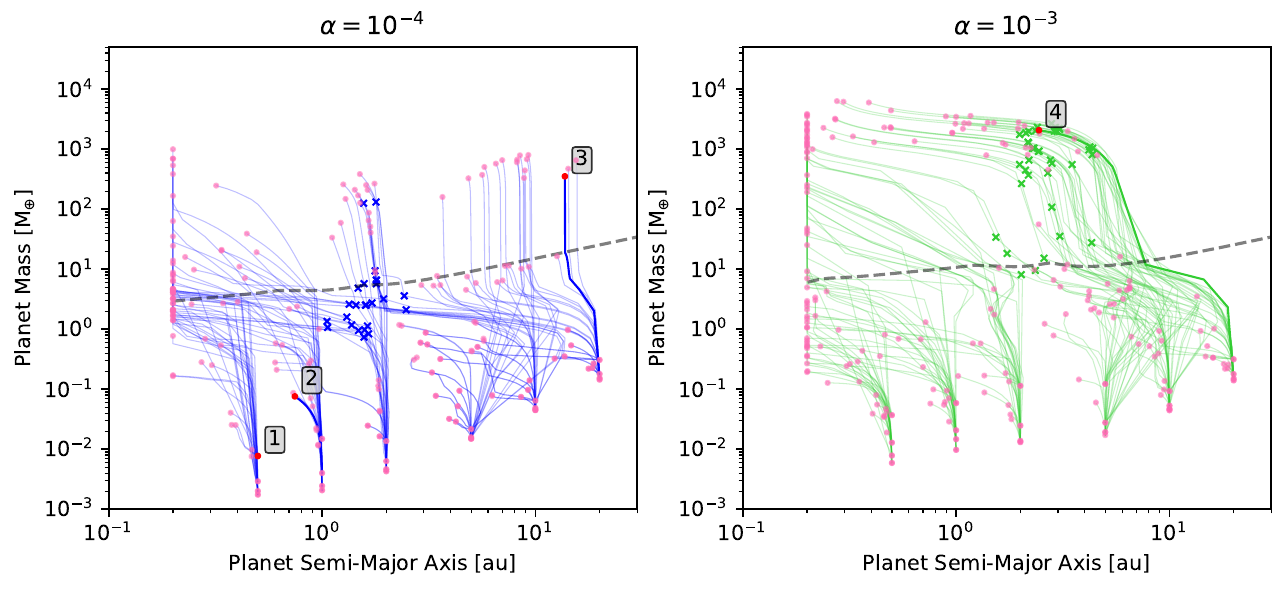}
    \caption{Same as Fig.~\ref{fig:planet_growth_tracks} for with planet trajectories drawn only up to 1 Myr, the time at which the pebble fluxes reported in Fig.~\ref{fig:M_v_flux_1Myr} are evaluated.}\label{fig:planet_growth_tracks_1Myr}
\end{figure*}

For completeness, we show in Fig.~\ref{fig:planet_growth_tracks_1Myr} the planet mass and semi-major axis distribution (pink dots), and their growth tracks (blue and green), up to $t=1~\mathrm{Myr}$.

\bsp	
\label{lastpage}
\end{document}